\documentclass[12pt,titlepage]{article}
\usepackage{amsfonts}
\usepackage{times}
\usepackage{geometry}
\usepackage[doublespacing]{setspace}

\input{tcilatex}
\begin{document}

\title{Functional robust regression for longitudinal data}
\author{Daniel Gervini \\
\emph{Department of Mathematical Sciences}\\
\emph{University of Wisconsin--Milwaukee}}
\maketitle

\begin{abstract}
We present a robust regression estimator for longitudinal data, which is
especially suited for functional data that has been observed on sparse or
irregular time grids. We show by simulation that the proposed estimators
possess good outlier-resistance properties compared with the traditional
functional least-squares estimator. As an example of application, we study
the relationship between levels of oxides of nitrogen and ozone in the city
of San Francisco.

\end{abstract}

\section{Introduction}

In a typical longitudinal study, a number of variables are measured on a
group of individuals and the goal is to analyze the relationships between
the trajectories of the variables. In recent years, functional data analysis
has provided efficient ways to analyze longitudinal data. In many cases the
variable trajectories are discretized continuous curves that can be
reconstructed by smoothing, and functional linear regression methods can be
applied to study the relationship between the variables (Ramsay and
Silverman, 2005). But in other situations the data is observed at sparse and
irregular time points, which makes smoothing difficult or even unfeasible.
Therefore, functional regression methods that can be applied directly to the
raw measurements become very useful.

Methods for functional data analysis of irregularly sampled curves have been
proposed by a number of authors, for the one-sample problem as well as for
the functional regression problem (Chiou et al., 2004; James et al., 2000; M%
\"{u}ller et al., 2008; Yao et al., 2005a, 2005b). Outlier-resistant
techniques for the functional one-sample problem have also been proposed
(Cuevas et al., 2007; Gervini, 2008, 2009; Fraiman and Muniz, 2001;
Locantore et al., 1999), and two recent papers deal with robust functional
regression for pre-smoothed curves (Zhu et al.~2011; Maronna and Yohai,
2012). However, outlier-resistant functional regression methods for raw
functional data have not yet been proposed in the literature. In this paper
we address this problem and present a computationally simple approach based
on random-effect models. Our simulations show that this method attains the
desired outlier resistance against atypical curves, and that the asymptotic
distribution of the test statistic is approximately valid for small samples.

As an example of application, we will analyze the daily trajectories of
oxides of nitrogen and ozone levels in the city of Sacramento, California,
during the summer of 2005. The data is shown in Figure \ref%
{fig:sample_curves}. The goal is to predict ozone concentration from oxides
of nitrogen. Both types of curves follow regular patterns, but some atypical
curves can be discerned in the sample. We will show in Section \ref%
{sec:Example} that to a large extend it is indeed possible to predict ozone
levels from oxides-of-nitrogen levels, but that the outlying curves distort
the classical regression estimators and that the proposed robust method
gives more reliable results.

The paper is organized as follows. Section \ref{sec:Methods} presents a
brief overview of functional linear regression and introduces the new
method. Section \ref{sec:Simulations} reports the results of a comparative
simulation study, and Section \ref{sec:Example} presents a detailed analysis
of the above mentioned ozone dataset. Technical derivations and proofs are
left to the Appendix. Matlab\registered\ programs implementing these
procedures are available on the author's webpage.

\FRAME{ftbpFU}{5.9819in}{2.0634in}{0pt}{\Qcb{Ozone Example. Daily
trajectories of ground-level concentrations of (a) oxides of nitrogen and
(b) ozone in the city of Sacramento in the Summer of 2005.}}{\Qlb{%
fig:sample_curves}}{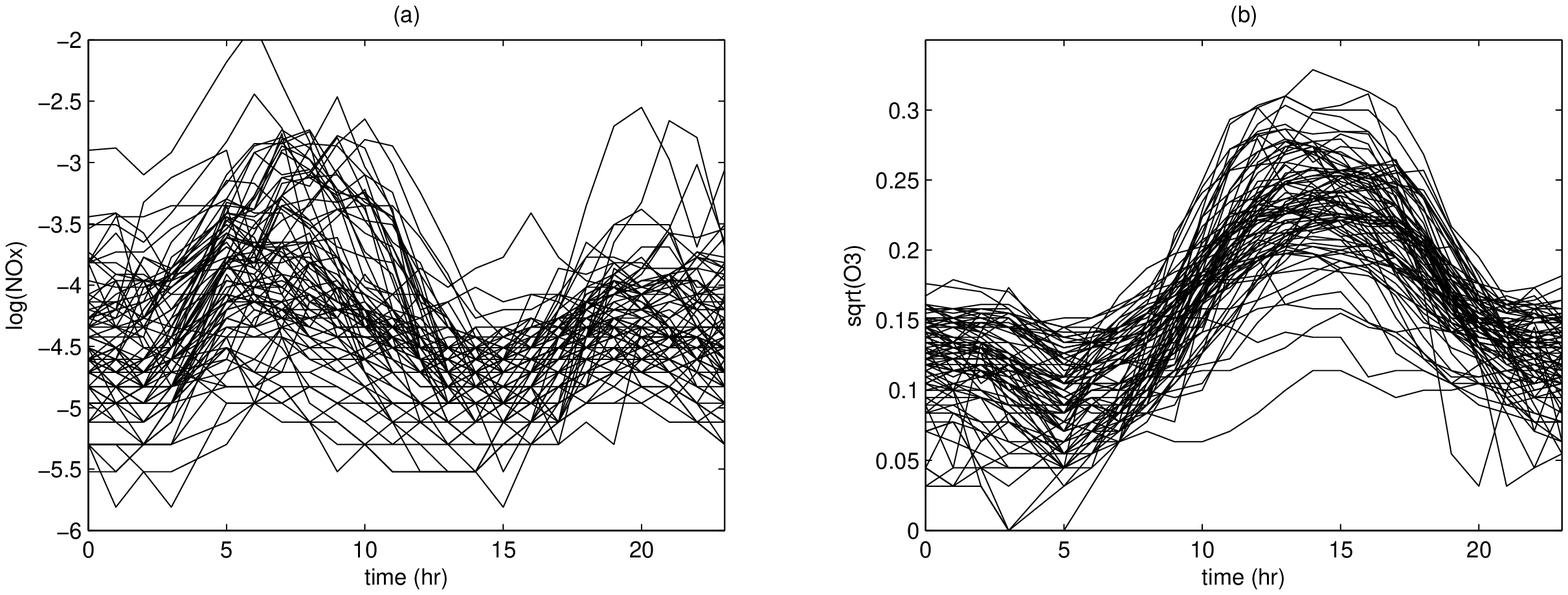}{\special{language "Scientific Word";type
"GRAPHIC";maintain-aspect-ratio TRUE;display "USEDEF";valid_file "F";width
5.9819in;height 2.0634in;depth 0pt;original-width 11.7372in;original-height
3.7014in;cropleft "0.0777";croptop "1";cropright "1";cropbottom "0";filename
'sample.eps';file-properties "XNPEU";}}

\section{\label{sec:Methods}Method}

\subsection{Background: classical functional linear regression}

The functional approach to longitudinal data analysis assumes that the
observations $(\mathbf{x}_{1},\mathbf{y}_{1}),\ldots ,\allowbreak (\mathbf{x}%
_{n},\mathbf{y}_{n})$ are discrete measurements of underlying continuous
curves, so 
\begin{eqnarray}
x_{ij} &=&X_{i}(s_{ij})+\varepsilon _{ij},\ \ i=1,\ldots ,n,\ \ j=1,\ldots
,m_{i},  \label{eq:raw-x} \\
y_{ij} &=&Y_{i}(t_{ij})+\varepsilon _{ij}^{\prime },\ \ i=1,\ldots ,n,\ \
j=1,\ldots ,m_{i}^{\prime },  \label{eq:raw-y}
\end{eqnarray}%
where $\{X_{i}(s)\}$ and $\{Y_{i}(t)\}$ are the trajectories of interest, $%
\{\varepsilon _{ij}\}$ and $\{\varepsilon _{ij}^{\prime }\}$ are random
measurement errors, and $\{s_{ij}\}$ and $\{t_{ij}\}$ are the time points
where the data is observed. The $X_{i}(s)$s and the $Y_{i}(t)$s are random
functions that we assume independent and identically distributed
realizations of a pair $(X(s),Y(t))$.

Suppose $X(s)$ and $Y(t)$ are square-integrable functions on an interval $%
[a,b]$. Define the norm $\Vert f\Vert =\{\int_{a}^{b}f^{2}(s)ds\}^{1/2}$ and
the inner product $\langle f,g\rangle =\int_{a}^{b}f(s)g(s)ds$. If $\mathrm{E%
}(\Vert X\Vert ^{2})$ and $\mathrm{E}(\Vert Y\Vert ^{2})$ are finite, then $%
X(s)$ and $Y(t)$ admit the decomposition 
\begin{eqnarray}
X(s) &=&\mu _{X}(s)+\sum_{k=1}^{p}U_{k}\phi _{k}(s),  \label{eq:KL-X} \\
Y(t) &=&\mu _{Y}(t)+\sum_{l=1}^{q}V_{l}\psi _{l}(t),  \label{eq:KL-Y}
\end{eqnarray}%
known as the Karhunen--Lo\`{e}ve decomposition (Ash and Gardner 1975,
ch.~1.4), where $\mu _{X}(s)=\mathrm{E}\{X(s)\}$, $\mu _{Y}(t)=\mathrm{E}%
\{Y(t)\}$, $\{\phi _{k}(s)\}$ and $\{\psi _{l}(t)\}$ are orthonormal
functions (i.e.$~\langle \phi _{k},\phi _{k^{\prime }}\rangle =\delta
_{kk^{\prime }}$ and $\langle \psi _{l},\psi _{l^{\prime }}\rangle =\delta
_{ll^{\prime }}$, where $\delta $ is Kronecker's delta), and $\{U_{k}\}$ and 
$\{V_{l}\}$ are random variables with zero mean and finite variance (without
loss of generality, one can assume that $\mathrm{var}(U_{1})\geq \mathrm{var}%
(U_{2})\geq \cdots >0$ and $\mathrm{var}(V_{1})\geq \mathrm{var}(V_{2})\geq
\cdots >0$.) This is the functional equivalent of the principal-component
decomposition in multivariate analysis, so the $\phi _{k}(s)$s and $\psi
_{l}(t)$s are called \textquotedblleft principal
components\textquotedblright ,\ and the $U_{k}$s and $V_{l}$s are called
\textquotedblleft component scores\textquotedblright . In principle $p$ and $%
q$ in (\ref{eq:KL-X}) and (\ref{eq:KL-Y}) could be infinite, but since $%
\mathrm{E}(\Vert X-\mu _{X}\Vert ^{2})=\sum_{k=1}^{p}\mathrm{var}(U_{k})$
and $\mathrm{E}(\Vert Y-\mu _{Y}\Vert ^{2})=\sum_{l=1}^{q}\mathrm{var}%
(V_{l}) $ are finite, the sequences $\{\mathrm{var}(U_{k})\}$ and $\{\mathrm{%
var}(V_{l})\}$ usually decrease to zero fast enough that for practical
purposes $p$ and $q$ can be assumed to be finite.

Methods for estimating the mean and the principal components of $X(s)$ and $%
Y(t)$ can be found in Ramsay and Silverman (2005), James et al.~(2000), and
Yao et al.~(2005b). These methods are not resistant to outliers, though;
outlier-resistant estimators of the mean and principal components have been
proposed by Locantore et al.~(1999), Cuevas et al.~(2007), and Gervini
(2008, 2009). We will use the method of Gervini (2009) to estimate the mean
and the principal components in (\ref{eq:KL-X}) and (\ref{eq:KL-Y}). This
method is briefly reviewed in the Appendix.

Now suppose that there is a functional linear relationship between $X(s)$
and $Y(t)$: 
\begin{equation}
Y(t)=\alpha _{0}(t)+\int_{a}^{b}\beta _{0}(s,t)X(s)ds+Z(t),
\label{eq:linear_model}
\end{equation}%
where $\alpha _{0}(t)$ is the intercept, $\beta _{0}(s,t)$ the slope, and $%
Z(t)$ the error term. We assume $\mathrm{E}\{Z(t)\}=0$ and $\mathrm{cov}%
\{X(s),Z(t)\}=0$ for all $s$ and $t$. (Note that the $Z$ is not necessarily
white noise; it is just the portion of $Y$ that is not explained by $X$, and
it is usually a smooth non-trivial process.) Since (\ref{eq:linear_model})
implies that $\mu _{Y}(t)=\alpha _{0}(t)+\int_{a}^{b}\beta _{0}(s,t)\mu
_{X}(s)ds$, we can rewrite (\ref{eq:linear_model}) as 
\begin{equation}
Y(t)=\mu _{Y}(t)+\int_{a}^{b}\beta _{0}(s,t)\{X(s)-\mu _{X}(s)\}\mathrm{ds}%
+Z(t).  \label{eq:linear_model_2}
\end{equation}%
Then the only parameter that remains to be estimated is the regression slope 
$\beta _{0}$.

Since $\{\phi _{k}\}$ is an orthonormal basis of the $X$-space and $\{\psi
_{l}\}$ is an orthonormal basis of the $Y$-space, without loss of generality
the regression slope can be expressed as 
\begin{equation}
\beta _{0}(s,t)=\sum_{k=1}^{p}\sum_{l=1}^{q}\theta _{0kl}\phi _{k}(s)\psi
_{l}(t).  \label{eq:Beta}
\end{equation}%
In matrix form, $\beta _{0}(s,t)=\mathbf{\phi }(s)^{T}\mathbf{\Theta }_{0}%
\mathbf{\psi }(t)$, where $\mathbf{\phi }(s)=(\phi _{1}(s),\ldots ,\phi
_{p}(s))^{T}$ and $\mathbf{\psi }(t)=(\psi _{1}(t),\ldots ,\psi _{q}(t))^{T}$%
. If we also collect the component scores $\{U_{k}\}$ and $\{V_{l}\}$ into
vectors $\mathbf{U}\in \mathbb{R}^{p}$ and $\mathbf{V}\in \mathbb{R}^{q}$,
from (\ref{eq:KL-X}), (\ref{eq:KL-Y}), (\ref{eq:linear_model_2}) and (\ref%
{eq:Beta}) we obtain 
\begin{eqnarray*}
\mathbf{\psi }(t)^{T}\mathbf{V} &=&\int_{a}^{b}\mathbf{\psi }(t)^{T}\mathbf{%
\Theta }_{0}^{T}\mathbf{\phi }(s)\mathbf{\phi }(s)^{T}\mathbf{U\ }\mathrm{ds}%
+Z(t) \\
&=&\mathbf{\psi }(t)^{T}\mathbf{\Theta }_{0}^{T}\mathbf{U}+\mathbf{\psi }%
(t)^{T}\mathbf{W},
\end{eqnarray*}%
where $\mathbf{W}\in \mathbb{R}^{q}$ is the random vector with elements $%
W_{l}=\langle Z,\psi _{l}\rangle $. This reduces the functional regression
model (\ref{eq:linear_model_2}) to a simpler multivariate regression model, 
\begin{equation}
\mathbf{V}=\mathbf{\Theta }_{0}^{T}\mathbf{U}+\mathbf{W},
\label{eq:reduced_linear_model}
\end{equation}%
and the problem now is to estimate the regression matrix $\mathbf{\Theta }%
_{0}$.

\subsection{\label{sec:robust-estim}Outlier-resistant functional regression}

As explained above, given the data $(\mathbf{x}_{1},\mathbf{y}_{1}),\ldots
,\allowbreak (\mathbf{x}_{n},\mathbf{y}_{n})$ we use the reduced-rank $t$
estimators of Gervini (2009) to obtain robust estimators of $\mu _{X}$, $\mu
_{Y}$, $\{\phi _{k}\}$, $\{\psi _{l}\}$, $\{U_{ik}\}$ and $\{V_{il}\}$. By (%
\ref{eq:Beta}) and (\ref{eq:reduced_linear_model}), the least-squares
estimator of $\beta _{0}(s,t)$ would be $\mathbf{\phi }(s)^{T}\mathbf{\hat{%
\Theta}\psi }(t)$ with 
\begin{equation}
\mathbf{\hat{\Theta}}=\limfunc{argmin}\limits_{\mathbf{\Theta }%
}\sum_{i=1}^{n}\Vert \mathbf{\hat{V}}_{i}-\mathbf{\Theta }^{T}\mathbf{\hat{U}%
}_{i}\Vert ^{2}=(\sum_{i=1}^{n}\mathbf{\hat{U}}_{i}\mathbf{\hat{U}}%
_{i}^{T})^{-1}\sum_{i=1}^{n}\mathbf{\hat{U}}_{i}\mathbf{\hat{V}}_{i}^{T}.
\label{eq:LSE}
\end{equation}%
However, this estimator is not robust. Although the reduced-rank $t$
estimators of $\mu _{X}$, $\mu _{Y}$, $\{\phi _{k}\}$ and $\{\psi _{l}\}$
are robust, the component scores $\mathbf{\hat{U}}_{i}$ and $\mathbf{\hat{V}}%
_{i}$ are individual parameters that will be outliers if the corresponding
curves $X_{i}(s)$ and $Y_{i}(t)$ are outliers. Therefore, the estimator of $%
\mathbf{\Theta }_{0}$ has to incorporate a mechanism to downweight outlying $%
\mathbf{\hat{U}}_{i}$s and $\mathbf{\hat{V}}_{i}$s.

This can be accomplished, for instance, by a modification of the $t$-type
GM-estimators of He et al.~(2000), that we will call GMt for short. Let 
\begin{equation}
(\mathbf{\hat{\Theta}},\mathbf{\hat{\Sigma}})=\limfunc{argmin}\limits_{%
\mathbf{\Theta ,\Sigma }}\sum_{i=1}^{n}\rho \{w(\mathbf{\hat{U}}_{i})(%
\mathbf{\hat{V}}_{i}-\mathbf{\Theta }^{T}\mathbf{\hat{U}}_{i})^{T}\mathbf{%
\Sigma }^{-1}(\mathbf{\hat{V}}_{i}-\mathbf{\Theta }^{T}\mathbf{\hat{U}}%
_{i})\}+n\log \left\vert \mathbf{\Sigma }\right\vert ,  \label{eq:WLNE}
\end{equation}%
where $\rho (x)=(\nu +q)\log \left( 1+x/\nu \right) $. These are the maximum
likelihood estimators of $\mathbf{\Theta }_{0}$ and $\mathbf{\Sigma }_{0}$
when $\mathbf{W}$ in (\ref{eq:reduced_linear_model}) follows a multivariate $%
t$ distribution with mean zero and scatter matrix $\mathbf{\Sigma }_{0}/w(%
\mathbf{\hat{U}}_{i})$, although we do not actually assume that $\mathbf{W}$
follows this distribution; as in He et al.~(2000), this is just the
motivation behind definition (\ref{eq:WLNE}).

It is shown in the Appendix that $\mathbf{\hat{\Theta}}$ and $\mathbf{\hat{%
\Sigma}}$ satisfy the fixed-point equations 
\begin{eqnarray}
\mathbf{\hat{\Theta}} &\mathbf{=}&\left\{ \sum_{i=1}^{n}\rho ^{\prime
}(e_{i})w(\mathbf{\hat{U}}_{i})\mathbf{\hat{U}}_{i}\mathbf{\hat{U}}%
_{i}^{T}\right\} ^{-1}\sum_{i=1}^{n}\rho ^{\prime }(e_{i})w(\mathbf{\hat{U}}%
_{i})\mathbf{\hat{U}}_{i}\mathbf{\hat{V}}_{i}^{T},
\label{eq:fixed-point-Theta} \\
\mathbf{\hat{\Sigma}} &=&\frac{1}{n}\sum_{i=1}^{n}\rho ^{\prime }(e_{i})w(%
\mathbf{\hat{U}}_{i})\mathbf{R}_{i}\mathbf{R}_{i}^{T},
\label{eq:fixed-point-Sigma}
\end{eqnarray}%
where $\mathbf{R}_{i}=\mathbf{\hat{V}}_{i}-\mathbf{\hat{\Theta}}^{T}\mathbf{%
\hat{U}}_{i}$ and $e_{i}=w(\mathbf{\hat{U}}_{i})\mathbf{R}_{i}^{T}\mathbf{%
\hat{\Sigma}}^{-1}\mathbf{R}_{i}$. These equations can be solved iteratively
by a reweighting algorithm.

As for the weights $w(\mathbf{\hat{U}}_{i})$, they are essentially a
by-product of the estimation of $\mu _{X}$, $\{\phi _{k}\}$ and $\{\mathbf{U}%
_{i}\}$. Since $E(\mathbf{U}_{i})=\mathbf{0}$ and $\mathrm{var}(\mathbf{U}%
_{i})=\mathrm{diag}(\lambda _{1},\ldots ,\lambda _{p})$, the $\mathbf{\hat{U}%
}_{i}$s are approximately uncorrelated with mean zero. The squared
Mahalanobis distance of $\mathbf{\hat{U}}_{i}$ is then $D_{i}^{2}=%
\sum_{k=1}^{p}\hat{U}_{ik}^{2}/\hat{\lambda}_{k}$, and large $D_{i}^{2}$s
will correspond to $X$-outliers. The $D_{i}^{2}$s will follow an approximate 
$\chi _{p}^{2}$ distribution if the data is Gaussian.

This suggests a number of weighting schemes. One possibility is to use
\textquotedblleft metric\textquotedblright\ trimming, 
\begin{equation}
w(\mathbf{\hat{U}}_{i})=\left\{ 
\begin{array}{lll}
1, &  & D_{i}^{2}\leq \chi _{p,1-\alpha }^{2}, \\ 
0, &  & \text{otherwise,}%
\end{array}%
\right.  \label{eq:w-metric}
\end{equation}%
where $\chi _{p,1-\alpha }^{2}$ is the $1-\alpha $ quantile of the $\chi
_{p}^{2}$ distribution. Another possibility is to use rank-based\ trimming, 
\begin{equation}
w(\mathbf{\hat{U}}_{i})=\left\{ 
\begin{array}{lll}
1, &  & \mathrm{rank}(D_{i}^{2})/n\leq 1-\alpha , \\ 
0, &  & \text{otherwise.}%
\end{array}%
\right.  \label{eq:w-rank}
\end{equation}%
The latter will always eliminate the $\alpha n$ observations with largest
Mahalanobis distances, even if they are not actual outliers; so we recommend
not using an unnecessarily large $\alpha $ for rank-based trimming. In
practice, the choice of $\alpha $ can be based on the proportion of outliers
observed in a boxplot or histogram of the $D_{i}^{2}$s.

The estimator $\mathbf{\hat{\Theta}}$ defined above belongs to the general
class of M-estimators, which have well-known asymptotic properties (Van der
Vaart, 1998, ch.~5). As shown in the Appendix, $\sqrt{n}\{\mathrm{vec}(%
\mathbf{\hat{\Theta})-}\mathrm{vec}(\mathbf{\Theta }_{0}\mathbf{)}\}$
follows an approximate $\mathrm{N}(\mathbf{0},\mathbf{A}^{-1}\mathbf{BA}%
^{-1})$ distribution for large $n$, with 
\begin{eqnarray}
\mathbf{A} &=&2\mathrm{E}\left\{ \rho ^{\prime \prime }(e)w^{2}(\mathbf{U})%
\mathbf{\Sigma }_{0}^{-1}\mathbf{RR}^{T}\otimes \mathbf{UU}^{T}\right\} +%
\mathbf{I}_{q}\otimes \mathrm{E}\left\{ \rho ^{\prime }(e)w(\mathbf{U})%
\mathbf{UU}^{T}\right\} ,  \label{eq:A} \\
\mathbf{B} &=&\mathrm{E}\left[ \{\rho ^{\prime }(e)\}^{2}w^{2}(\mathbf{U})%
\mathbf{RR}^{T}\otimes \mathbf{UU}^{T}\right] .  \label{eq:B}
\end{eqnarray}%
The matrices $\mathbf{A}$ and $\mathbf{B}$ can be easily estimated,
replacing expectations by averages. This asymptotic distribution can be
used, for instance, to test significance of the regression: if $\mathbf{%
\Theta }_{0}=\mathbf{O}$, Wald's statistic $Q=n\mathrm{vec}(\mathbf{\hat{%
\Theta})}^{T}\mathbf{\hat{A}\hat{B}}^{-1}\mathbf{\hat{A}}\mathrm{vec}(%
\mathbf{\hat{\Theta})}$ follows an approximate $\chi _{pq}^{2}$ distribution
for large $n$, so we decide the regression is significant if $Q\geq \chi
_{pq,1-\alpha }^{2}$ for a given level $\alpha $. We can also construct
marginal tests and confidence intervals for the individual coefficients $%
\theta _{kl}$.

In Section \ref{sec:Simulations} we will study the accuracy of this
asymptotic approximation. It is our experience that the distribution of $%
\mathbf{\hat{\Theta}}$ approaches normality quite fast, but the above
\textquotedblleft sandwich formula\textquotedblright\ tends to underestimate
the variance when the sample size $n$ is small. In that case it is better to
use bootstrap estimators of the covariance matrix of $\mathrm{vec}(\mathbf{%
\hat{\Theta})}$.

\section{Simulations\label{sec:Simulations}}

In this section we study by simulation the finite-sample behavior of the
estimators (\ref{eq:WLNE}). To this end, we generated data from model (\ref%
{eq:reduced_linear_model}) with $\mathbf{U}\sim \mathrm{N}(\mathbf{0},%
\mathbf{\Lambda })$ and $\mathbf{W}\sim \mathrm{N}(\mathbf{0},\mathbf{\Sigma 
})$, where $\mathbf{\Lambda }=\mathrm{diag}(1,1/2,\ldots ,1/p)$ and $\mathbf{%
\Sigma }=\mathrm{diag}(1,1/2,\ldots ,1/q)$. Two regression parameters $%
\mathbf{\Theta }_{0}$ were considered: for the first set of simulations (to
study estimation error) we took $\mathbf{\Theta }_{0}$ with $\theta
_{0,11}=3 $ and $\theta _{0,ij}=0$ for $(i,j)\neq (1,1)$; for the second set
of simulations (to study the goodness of the asymptotic approximation of
Wald's test) we took $\mathbf{\Theta }_{0}=\mathbf{O}$. The curves $%
\{X_{i}(s)\}$ and $\{Y_{i}(t)\}$ were generated following (\ref{eq:KL-X})
and (\ref{eq:KL-Y}), with $\mu _{X}(s)$ and $\mu _{Y}(t)$ equal to zero, $%
\phi _{k}(s)=\sqrt{2}\sin (k\pi s)$ and $\psi _{l}(t)=\sqrt{2}\sin (l\pi t)$%
, for $s$ and $t$ in $[0,1]$. The raw observations were generated following (%
\ref{eq:raw-x}) and (\ref{eq:raw-y}), with random $s_{ij}$s uniformly
distributed in $[0,1]$, $\{\varepsilon _{ij}\}$ and $\{\varepsilon
_{ij}^{\prime }\}$ independent $\mathrm{N}(0,0.01)$, and $%
m_{i}=m_{i}^{\prime }=m$; for simplicity we took the grid $\{t_{ij}\}$ equal
to $\{s_{ij}\}$.

The first series of simulations were designed to study estimation error of
the $\mathbf{\hat{\Theta}}$s, both for clean and for outlier-contaminated
data. We generated outliers by replacing $[\varepsilon n]$ of the pairs $(%
\mathbf{U}_{i},\mathbf{V}_{i})$ by $(\mathbf{U}_{i}^{\ast },\mathbf{V}%
_{i}^{\ast })$, with $U_{i1}^{\ast }=U_{i1}+5$ and $U_{ij}^{\ast }=U_{ij}$
for $j\neq 1$, and $\mathbf{V}_{i}^{\ast }=\mathbf{W}_{i}$. Note that the
contaminated data $(\mathbf{U}_{i}^{\ast },\mathbf{V}_{i}^{\ast })$ follows
model (\ref{eq:reduced_linear_model}) with $\mathbf{\Theta }_{0}=\mathbf{O}$
and high-leverage $\mathbf{U}_{i}^{\ast }$s, so the effect of this type of
contamination is an underestimation of $\theta _{0,11}$ that tends to pull $%
\hat{\beta}(s,t)$ towards 0.

The estimation of $\mathbf{\Theta }_{0}$ requires two steps: first, to
estimate $\{\mathbf{U}_{i}\}$ and $\{\mathbf{V}_{i}\}$ from the raw data,
and then to compute $\mathbf{\hat{\Theta}}$ from the $\mathbf{\hat{U}}_{i}$s
and the $\mathbf{\hat{V}}_{i}$s. So we compared two procedures: a non-robust
procedure, using reduced-rank Normal models (James et al., 2000) to estimate
the component scores, followed by the ordinary least-squares regression
estimator (\ref{eq:LSE}); and a robust procedure, using reduced-rank $t$%
-models (Gervini, 2009) to estimate the component scores, followed by the
GMt regression estimator (\ref{eq:WLNE}). For the robust procedure, we
considered the two types of weights $w(\mathbf{\hat{U}}_{i})$ discussed in
Section \ref{sec:robust-estim}, with trimming proportions $\alpha =.10$ and $%
\alpha =.50$; degrees of freedom $\nu =1$ and $\nu =5$ were used for the $t$%
-models.

Four levels of contamination $\varepsilon $ were considered: 0 (clean data), 
$.10$, $.20$ and $.30$. We took $n=50$ as sample size, $m=20$ as grid size,
and $p=q=2$ as model dimensions. Each case was replicated 1000 times. As
measure of the estimation error we used the expected root integrated squared
error $\mathrm{E}(\Vert \hat{\beta}-\beta _{0}\Vert )$, where $\Vert \hat{%
\beta}-\beta _{0}\Vert ^{2}=\int_{0}^{1}\int_{0}^{1}\{\hat{\beta}(s,t)-\beta
_{0}(s,t)\}^{2}\ ds\ dt$.

The results are reported in Table \ref{tab:Simulations_1}, along with Monte
Carlo standard errors. We see that for non-contaminated data ($\varepsilon
=0 $), there is no significant difference between metric and rank trimming
for a given pair $(\nu ,\alpha )$. The trimming proportion $\alpha $ has a
larger impact on the estimator's behavior than the degrees of freedom $\nu $%
. For this reason we recommend choosing $\alpha $ adaptively, so as not to
cut off too much good data. When $\varepsilon >0$, we see that metric
trimming\ tends to outperform rank trimming\ for a given pair $(\nu ,\alpha
) $. Somewhat counterintuitively, estimators with $\nu =5$ tend to be more
robust than those with $\nu =1$ for a given $\alpha $; the reason is that
for this type of contamination, which affects $\mathbf{\hat{\Theta}}$ but
not the $\hat{\phi}_{k}$s or the $\hat{\psi}_{l}$s, $t$ models with $\nu =5$
provide more accurate estimators of $\{\mathbf{U}_{i}\}$ and $\{\mathbf{V}%
_{i}\}$ than $t$ models with $\nu =1$ (for other types of contamination this
is no longer true, although $t$ models with $\nu =5$ are still very robust;
see Gervini (2009).) In general, then, the recommendation is to use $t$%
-model estimators with metrically trimmed weights and a trimming proportion
chosen adaptively.

\begin{table}[tbp] \centering%
\begin{tabular}{llcccc}
\hline
&  & \multicolumn{4}{c}{Contamination proportion} \\ \cline{3-6}
\multicolumn{1}{c}{Estimator} & \multicolumn{1}{c}{} & 0\% & 10\% & 20\% & 
30\% \\ \hline
&  & \multicolumn{1}{l}{} & \multicolumn{1}{l}{} & \multicolumn{1}{l}{} & 
\multicolumn{1}{l}{} \\ 
\multicolumn{1}{r}{Least squares} &  & \multicolumn{1}{l}{.293 (.004)} & 
\multicolumn{1}{l}{2.241 (.006)} & \multicolumn{1}{l}{2.644 (.048)} & 
\multicolumn{1}{l}{2.731 (.007)} \\ 
&  & \multicolumn{1}{l}{} & \multicolumn{1}{l}{} & \multicolumn{1}{l}{} & 
\multicolumn{1}{l}{} \\ 
GMt, $\nu =1$, $\alpha =.10$ &  & \multicolumn{1}{l}{} & \multicolumn{1}{l}{}
& \multicolumn{1}{l}{} & \multicolumn{1}{l}{} \\ 
\multicolumn{1}{r}{Metric trim} &  & \multicolumn{1}{l}{.472 (.006)} & 
\multicolumn{1}{l}{.497 (.007)} & \multicolumn{1}{l}{1.316 (.028)} & 
\multicolumn{1}{l}{2.924 (.008)} \\ 
\multicolumn{1}{r}{Rank trim} &  & \multicolumn{1}{l}{.473 (.006)} & 
\multicolumn{1}{l}{.469 (.007)} & \multicolumn{1}{l}{2.246 (.028)} & 
\multicolumn{1}{l}{2.941 (.006)} \\ 
&  & \multicolumn{1}{l}{} & \multicolumn{1}{l}{} & \multicolumn{1}{l}{} & 
\multicolumn{1}{l}{} \\ 
GMt, $\nu =1$, $\alpha =.50$ &  & \multicolumn{1}{l}{} & \multicolumn{1}{l}{}
& \multicolumn{1}{l}{} & \multicolumn{1}{l}{} \\ 
\multicolumn{1}{r}{Metric trim} &  & \multicolumn{1}{l}{.846 (.012)} & 
\multicolumn{1}{l}{.800 (.013)} & \multicolumn{1}{l}{1.112 (.018)} & 
\multicolumn{1}{l}{1.756 (.022)} \\ 
\multicolumn{1}{r}{Rank trim} &  & \multicolumn{1}{l}{.832 (.012)} & 
\multicolumn{1}{l}{.922 (.015)} & \multicolumn{1}{l}{1.212 (.018)} & 
\multicolumn{1}{l}{1.784 (.021)} \\ 
&  & \multicolumn{1}{l}{} & \multicolumn{1}{l}{} & \multicolumn{1}{l}{} & 
\multicolumn{1}{l}{} \\ 
GMt, $\nu =5$, $\alpha =.10$ &  & \multicolumn{1}{l}{} & \multicolumn{1}{l}{}
& \multicolumn{1}{l}{} & \multicolumn{1}{l}{} \\ 
\multicolumn{1}{r}{Metric trim} &  & \multicolumn{1}{l}{.379 (.005)} & 
\multicolumn{1}{l}{.396 (.005)} & \multicolumn{1}{l}{1.493 (.023)} & 
\multicolumn{1}{l}{2.746 (.006)} \\ 
\multicolumn{1}{r}{Rank trim} &  & \multicolumn{1}{l}{.374 (.005)} & 
\multicolumn{1}{l}{.395 (.006)} & \multicolumn{1}{l}{2.341 (.011)} & 
\multicolumn{1}{l}{2.792 (.005)} \\ 
&  & \multicolumn{1}{l}{} & \multicolumn{1}{l}{} & \multicolumn{1}{l}{} & 
\multicolumn{1}{l}{} \\ 
GMt, $\nu =5$, $\alpha =.50$ &  & \multicolumn{1}{l}{} & \multicolumn{1}{l}{}
& \multicolumn{1}{l}{} & \multicolumn{1}{l}{} \\ 
\multicolumn{1}{r}{Metric trim} &  & \multicolumn{1}{l}{.795 (.010)} & 
\multicolumn{1}{l}{.666 (.011)} & \multicolumn{1}{l}{.912 (.015)} & 
\multicolumn{1}{l}{1.494 (.021)} \\ 
\multicolumn{1}{r}{Rank trim} &  & \multicolumn{1}{l}{.783 (.010)} & 
\multicolumn{1}{l}{.829 (.013)} & \multicolumn{1}{l}{1.054 (.017)} & 
\multicolumn{1}{l}{1.506 (.021)} \\ \hline
\end{tabular}%
\caption{Simulation Results. Mean root integrated squared errors of
$\hat{\beta}$ under various contamination proportions (Monte Carlo standard errors in parenthesis).}%
\label{tab:Simulations_1}%
\end{table}%

The second series of simulations were designed to assess the finite-sample
adequacy of the asymptotic Wald test. To this end we generated data as
before, but with $\mathbf{\Theta }_{0}=\mathbf{O}$. Then $Q=n\mathrm{vec}(%
\mathbf{\hat{\Theta})}^{T}\mathbf{\hat{\Omega}}^{-1}\mathrm{vec}(\mathbf{%
\hat{\Theta})}$ should approximately follow a $\chi _{pq}^{2}$ distribution,
where $\mathbf{\Omega }$ is the asymptotic covariance matrix of $\sqrt{n}%
\mathrm{vec}(\mathbf{\hat{\Theta})}$. For GMt estimators, $\mathbf{\Omega }$
is the \textquotedblleft sandwich formula\textquotedblright\ given in
Section \ref{sec:robust-estim}; for the least-squares estimator, $\mathbf{%
\Omega }=\mathrm{E}(\mathbf{RR}^{T})\otimes \{\mathrm{E}(\mathbf{UU}%
^{T})\}^{-1}$. Table \ref{tab:Simulations_2} reports the tail probabilities $%
\mathrm{P}(Q\geq \chi _{pq,1-\alpha }^{2})$ for the usual values of $\alpha $
($.10$, $.05$ and $.01$) and various combinations of parameters $n$, $m$, $p$
and $q$. Each combination was replicated 10,000 times. We compared only two
estimators this time: the least-squares estimator and the 10\% metrically
trimmed GMt estimator with $\nu =5$. We see in Table \ref{tab:Simulations_2}
that the asymptotic $\chi _{pq}^{2}$ approximation works reasonably well for
the least-squares estimator if the ratio $n/pq$ exceeds 15; however, for the
GMt estimator a ratio $n/pq$ of at least 35 is necessary for the asymptotic
approximation to be reasonably good. Therefore, the asymptotic Wald test can
be used with confidence only for large sample sizes and relatively small
dimensions. In other cases, permutation tests or Wald tests with
bootstrap-estimated covariances are preferable.

\begin{table}[tbp] \centering%
\begin{tabular}{llccc}
\hline
&  & \multicolumn{3}{c}{Nominal probability} \\ \cline{3-5}
\multicolumn{1}{c}{Parameters} & \multicolumn{1}{c}{Estimator} & .10 & .05 & 
.01 \\ \hline
&  & \multicolumn{1}{l}{} & \multicolumn{1}{l}{} & \multicolumn{1}{l}{} \\ 
$n=50$, $m=20$, & LS & \multicolumn{1}{l}{.1426 (.0035)} & 
\multicolumn{1}{l}{.0819 (.0027)} & \multicolumn{1}{l}{.0219 (.0015)} \\ 
$p=q=2$ & GMt & \multicolumn{1}{l}{.2270 (.0042)} & \multicolumn{1}{l}{.1571
(.0036)} & \multicolumn{1}{l}{.0749 (.0026)} \\ 
&  & \multicolumn{1}{l}{} & \multicolumn{1}{l}{} & \multicolumn{1}{l}{} \\ 
$n=100$, $m=20$, & LS & \multicolumn{1}{l}{.1272 (.0033)} & 
\multicolumn{1}{l}{.0693 (.0025)} & \multicolumn{1}{l}{.0170 (.0013)} \\ 
$p=q=2$ & GMt & \multicolumn{1}{l}{.1584 (.0037)} & \multicolumn{1}{l}{.0952
(.0029)} & \multicolumn{1}{l}{.0366 (.0019)} \\ 
&  & \multicolumn{1}{l}{} & \multicolumn{1}{l}{} & \multicolumn{1}{l}{} \\ 
$n=150$, $m=10$, & LS & \multicolumn{1}{l}{.1117 (.0032)} & 
\multicolumn{1}{l}{.0561 (.0023)} & \multicolumn{1}{l}{.0123 (.0011)} \\ 
$p=q=2$ & GMt & \multicolumn{1}{l}{.1392 (.0035)} & \multicolumn{1}{l}{.0824
(.0027)} & \multicolumn{1}{l}{.0258 (.0016)} \\ 
&  & \multicolumn{1}{l}{} & \multicolumn{1}{l}{} & \multicolumn{1}{l}{} \\ 
$n=100$, $m=20$, & LS & \multicolumn{1}{l}{.1452 (.0035)} & 
\multicolumn{1}{l}{.0813 (.0027)} & \multicolumn{1}{l}{.0211 (.0014)} \\ 
$p=q=3$ & GMt & \multicolumn{1}{l}{.2750 (.0045)} & \multicolumn{1}{l}{.1900
(.0039)} & \multicolumn{1}{l}{.0875 (.0028)} \\ 
&  & \multicolumn{1}{l}{} & \multicolumn{1}{l}{} & \multicolumn{1}{l}{} \\ 
$n=150$, $m=20$, & LS & \multicolumn{1}{l}{.1316 (.0034)} & 
\multicolumn{1}{l}{.0718 (.0026)} & \multicolumn{1}{l}{.0144 (.0012)} \\ 
$p=q=3$ & GMt & \multicolumn{1}{l}{.2111 (.0041)} & \multicolumn{1}{l}{.1360
(.0034)} & \multicolumn{1}{l}{.0514 (.0022)} \\ 
&  & \multicolumn{1}{l}{} & \multicolumn{1}{l}{} & \multicolumn{1}{l}{} \\ 
$n=200$, $m=10$, & LS & \multicolumn{1}{l}{.1185 (.0032)} & 
\multicolumn{1}{l}{.0625 (.0024)} & \multicolumn{1}{l}{.0169 (.0013)} \\ 
$p=q=3$ & GMt & \multicolumn{1}{l}{.1782 (.0038)} & \multicolumn{1}{l}{.1122
(.0032)} & \multicolumn{1}{l}{.0391 (.0019)} \\ \hline
\end{tabular}%
\caption{Simulation Results. Finite-sample tail probabilities of Wald's significance-of-regression test 
for nominal asymptotic probabilities .10, .05 and .01 (Monte Carlo standard errors in parenthesis).}%
\label{tab:Simulations_2}%
\end{table}%

\section{\label{sec:Example}Application: Ozone Pollution Data}

Ground-level ozone is an air pollutant known to cause serious health
problems. Unlike other pollutants, ozone is not emitted directly into the
air but forms as a result of complex chemical reactions, including volatile
organic compounds and oxides of nitrogen among other factors. Modeling
ground-level ozone formation has been an active topic of air-quality studies
for many years. The California Environmental Protection Agency database,
available at http://www.arb.ca.gov/aqd/aqdcd/aqdcddld.htm, has collected
data on hourly concentrations of pollutants at different locations in
California for the years 1980 to 2009. Here we will focus on the
trajectories of oxides of nitrogen (NOx) and ozone (O3) in the city of
Sacramento (site 3011 in the database) between June 6 and August 26 of 2005,
which make a total of 82 days (shown in Figure \ref{fig:sample_curves}).
There are a few days with some missing observations (9 in total), but since
the method can handle unequal time grids, imputation of the missing data was
not necessary.

\FRAME{ftbpFU}{5.3125in}{6.9427in}{0pt}{\Qcb{Ozone Example. Normal ($---$)
and Cauchy (-----) reduced-rank B-spline estimators of the mean [(a),(b)],
the first principal component [(c),(d)], the second principal component
[(e),(f)] and the third principal component [(g),(h)] of log-NOx and root-O3
trajectories. }}{\Qlb{fig:mean-pc}}{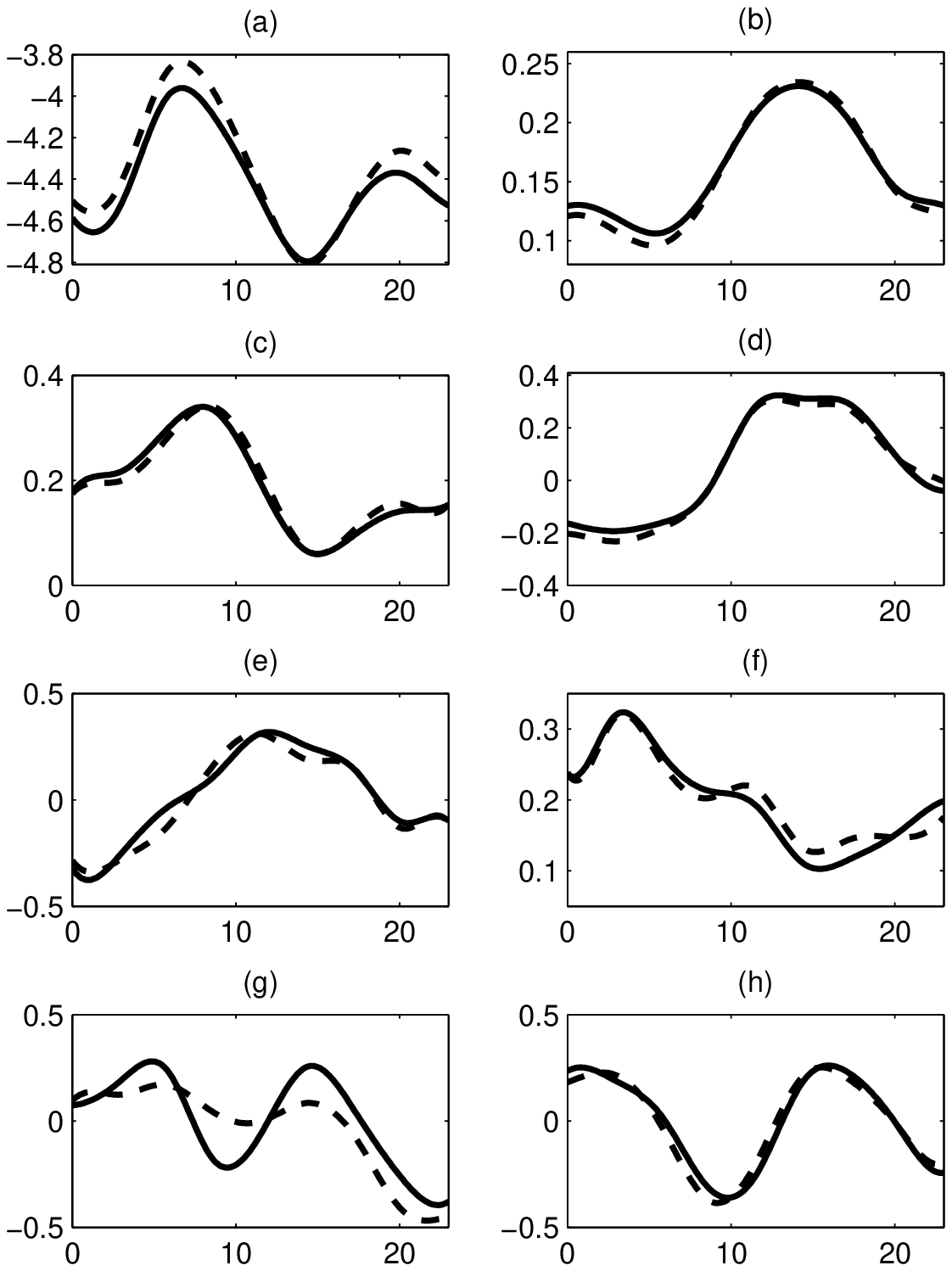}{\special{language
"Scientific Word";type "GRAPHIC";maintain-aspect-ratio TRUE;display
"USEDEF";valid_file "F";width 5.3125in;height 6.9427in;depth
0pt;original-width 5.2857in;original-height 6.915in;cropleft "0";croptop
"1";cropright "1";cropbottom "0";filename 'mean-pc.eps';file-properties
"XNPEU";}}

The first step in the analysis is to fit reduced-rank models to the sample
curves. We used cubic B-splines with 7 equally spaced knots every 5 years,
and fitted Normal and $t_{1}$ (Cauchy) reduced-rank models with up to 10
principal components. For both the response and the explanatory curves, the
leading three components explain at least 85\% of the total variability, so
we retained these models. The means and the principal components are plotted
in Figure \ref{fig:mean-pc}. There is no substantial difference between the
estimators obtained by these models, except perhaps for the mean and the
third component of log-NOx (Figures \ref{fig:mean-pc} (a) and (g)).

With the Normal component scores we computed the Least Squares estimator,
obtaining 
\[
\mathbf{\hat{\Theta}}_{LS}=\left( 
\begin{array}{ccc}
.0404 & -.0077 & .0083 \\ 
-.0537 & -.0085 & .0317 \\ 
-.0109 & -.0173 & -.0263%
\end{array}%
\right) . 
\]%
With the Cauchy component scores we computed the GMt estimator with 1 degree
of freedom and 10\% metric trimming, obtaining 
\[
\mathbf{\hat{\Theta}}_{GM}=\left( 
\begin{array}{ccc}
.0406 & -.0172 & .0045 \\ 
-.0451 & .0029 & .0266 \\ 
-.0289 & -.0071 & -.0317%
\end{array}%
\right) . 
\]%
The latter cut off 5 observations out of the 82. There are some noticeable
differences between these two estimators, even leaving aside the third row
(which are not easily comparable, since $\hat{\phi}_{LS,3}(s)$ and $\hat{\phi%
}_{GM,3}(s)$ are rather different). The differences are more striking in the
slope estimators $\hat{\beta}_{LS}(s,t)$ and $\hat{\beta}_{GM}(s,t)$, shown
in Figure \ref{fig:betas}. There is a \textquotedblleft
bump\textquotedblright\ in $\hat{\beta}_{GM}(s,t)$ around $(s,t)=(8,16)$
that does not appear in $\hat{\beta}_{LS}(s,t)$. This means that the robust
slope estimator assigns positive weight to NOx values around 8am in the
prediction of O3 levels around 4pm, showing that there is a persistent
effect of oxides-of-nitrogen level in ozone formation.

\FRAME{ftbpFU}{6.5094in}{2.3938in}{0pt}{\Qcb{Ozone Example. Functional slope
estimators obtained by (a) least squares using Normal scores and (b)
metric-trimmed GMt using Cauchy scores.}}{\Qlb{fig:betas}}{betas.tif}{%
\special{language "Scientific Word";type "GRAPHIC";maintain-aspect-ratio
TRUE;display "USEDEF";valid_file "F";width 6.5094in;height 2.3938in;depth
0pt;original-width 12.698in;original-height 4.3024in;cropleft
"0.0718";croptop "1";cropright "1";cropbottom "0";filename
'betas.tif';file-properties "XNPEU";}}

Of course, none of this would be meaningful if the regression model was not
statistically significant. But the estimated response curves, shown in
Figure \ref{fig:pred-y}, clearly show that the model does predict the
response curves to a large extent. The robust estimator provides a better
fit overall, with a root median squared error of $.022$ compared to the root
median squared error of $.023$ for the least squares estimator.

\FRAME{ftbpFU}{7.6104in}{1.881in}{0pt}{\Qcb{Ozone Example. Daily
trajectories of root-O3 levels: (a) observed, (b) predicted by robust GMt
estimator, and (c) predicted by least squares.}}{\Qlb{fig:pred-y}}{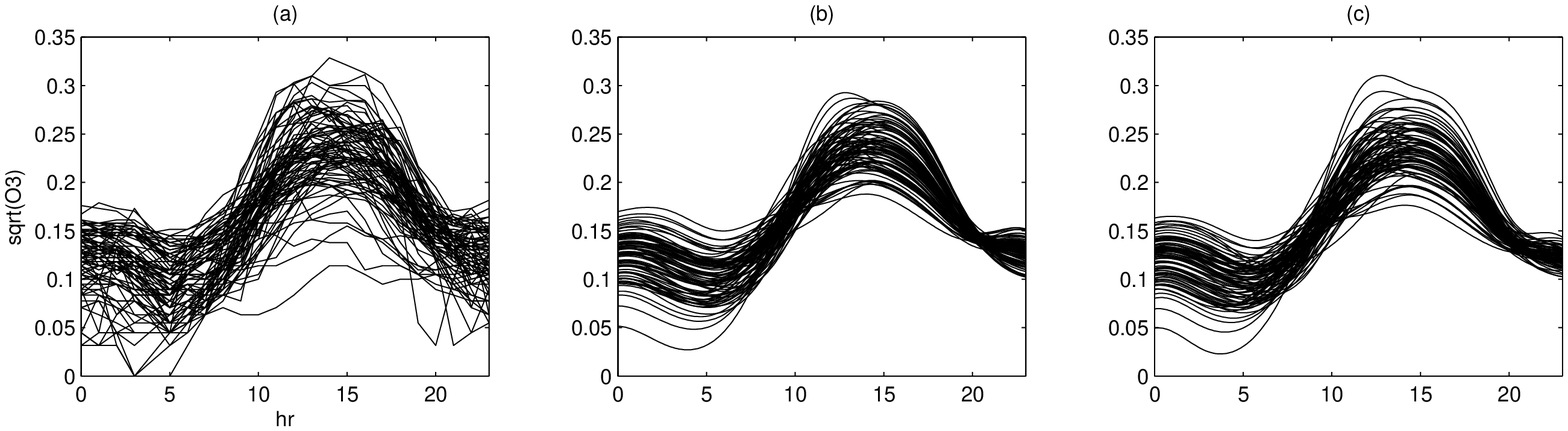%
}{\special{language "Scientific Word";type "GRAPHIC";maintain-aspect-ratio
TRUE;display "USEDEF";valid_file "F";width 7.6104in;height 1.881in;depth
0pt;original-width 12.6816in;original-height 2.8366in;cropleft
"0.0848";croptop "1";cropright "1";cropbottom "0";filename
'pred-y.eps';file-properties "XNPEU";}}

\section*{Acknowledgement}

The author was partly supported by NSF grants DMS 0604396 and 1006281.

\section*{Appendix}

\subsection*{Reduced-rank $t$ models}

The method proposed by Gervini (2009) to estimate the mean and the principal
components of a stochastic process $X$ works as follows. The mean function $%
\mu _{X}$ and the principal components $\{\phi _{k}\}$ are modeled as spline
functions; that is, given a set of spline basis functions $b_{1},\ldots
,b_{N}$, chosen by the user, it is assumed that $\mu
_{X}(s)=\sum_{l=1}^{N}\xi _{l}b_{l}(s)$ and $\phi _{k}(s)=\sum_{l=1}^{N}\eta
_{kl}b_{l}(s)$. The observed vector $\mathbf{x}_{i}$ can then be expressed
as 
\[
\mathbf{x}_{i}=\mathbf{B}_{i}\mathbf{\xi }+\mathbf{B}_{i}\mathbf{H\Lambda }%
^{1/2}\mathbf{z}_{i}+\sigma \mathbf{\varepsilon }_{i}, 
\]%
where $\mathbf{B}_{i}=[b_{l}(s_{ij})]_{(j,l)}$, $\mathbf{H}=[\mathbf{\eta }%
_{1},\ldots ,\mathbf{\eta }_{p}]$ and $\mathbf{\Lambda }=\mathrm{diag}%
(\lambda _{1},\ldots ,\lambda _{p})$. Note that $\mathbf{U}_{i}=\mathbf{%
\Lambda }^{1/2}\mathbf{z}_{i}$ in this notation. By assuming $(\mathbf{z}%
_{i},\mathbf{\varepsilon }_{i})$ has a standard multivariate $t$
distribution, robust maximum likelihood estimators of $\mathbf{\xi }$, $\{%
\mathbf{\eta }_{k}\}$, $\{\lambda _{k}\}$ and $\sigma $ are obtained. The
estimators are computed via a standard EM algorithm. The optimal number of
components $p$ can be chosen via AIC or BIC criteria. See Gervini (2009) for
details. In addition to parameter estimates, the EM algorithm yields
predictors of the random effects $\mathbf{z}_{i}$, so one obtains $\mathbf{%
\hat{U}}_{i}=\mathbf{\hat{\Lambda}}^{1/2}\mathbf{\hat{z}}_{i}$ as a
by-product. The estimators of $\mu _{Y}$, $\{\psi _{k}\}$, and $\{\mathbf{V}%
_{i}\}$ are obtained in a similar way from the sample $\mathbf{y}_{1},\ldots
,\mathbf{y}_{n}$.

\subsection*{GMt estimating equations and asymptotics}

The estimators $\mathbf{\hat{\Theta}}$ and $\mathbf{\hat{\Sigma}}$ defined
by (\ref{eq:WLNE}) are M-type estimators (Van der Vaart, 1998, ch.~5), since
they minimize a function of the form $M(\mathbf{\Theta },\mathbf{\Sigma })=%
\frac{1}{n}\sum_{i=1}^{n}m_{(\mathbf{\Theta },\mathbf{\Sigma })}(\mathbf{%
\hat{U}}_{i},\mathbf{\hat{V}}_{i})$. Specifically, 
\[
m_{(\mathbf{\Theta },\mathbf{\Sigma })}(\mathbf{\hat{U}}_{i},\mathbf{\hat{V}}%
_{i})=\rho \{w(\mathbf{\hat{U}}_{i})(\mathbf{\hat{V}}_{i}-\mathbf{\Theta }%
^{T}\mathbf{\hat{U}}_{i})^{T}\mathbf{\Sigma }^{-1}(\mathbf{\hat{V}}_{i}-%
\mathbf{\Theta }^{T}\mathbf{\hat{U}}_{i})\}+\log \left\vert \mathbf{\Sigma }%
\right\vert . 
\]%
Then $\mathbf{\hat{\Theta}}$ and $\mathbf{\hat{\Sigma}}$ solve the equations 
$\frac{\partial }{\partial \mathbf{\Theta }}M(\mathbf{\hat{\Theta}},\mathbf{%
\hat{\Sigma}})=\mathbf{O}$ and $\frac{\partial }{\partial \mathbf{\Sigma }}M(%
\mathbf{\hat{\Theta}},\mathbf{\hat{\Sigma}})=\mathbf{O}$. To compute matrix
derivatives we use the method of differentials (Magnus and Neudecker, 1999).
Differentiating with respect to $\mathbf{\Theta }$ we obtain 
\begin{eqnarray*}
\mathrm{d}m_{(\mathbf{\Theta },\mathbf{\Sigma })}(\mathbf{\hat{U}}_{i},%
\mathbf{\hat{V}}_{i}) &=&\rho ^{\prime }(e_{i})w(\mathbf{\hat{U}}_{i})2(%
\mathbf{\hat{V}}_{i}-\mathbf{\Theta }^{T}\mathbf{\hat{U}}_{i})^{T}\mathbf{%
\Sigma }^{-1}\{-(\mathrm{d}\mathbf{\Theta })^{T}\mathbf{\hat{U}}_{i}\} \\
&=&-2\rho ^{\prime }(e_{i})w(\mathbf{\hat{U}}_{i})\mathrm{tr}\{(\mathrm{d}%
\mathbf{\Theta })^{T}\mathbf{\hat{U}}_{i}(\mathbf{\hat{V}}_{i}-\mathbf{%
\Theta }^{T}\mathbf{\hat{U}}_{i})^{T}\mathbf{\Sigma }^{-1}\} \\
&=&-2\rho ^{\prime }(e_{i})w(\mathbf{\hat{U}}_{i})\mathrm{vec}(\mathrm{d}%
\mathbf{\Theta })^{T}\mathrm{vec}\{\mathbf{\hat{U}}_{i}(\mathbf{\hat{V}}_{i}-%
\mathbf{\Theta }^{T}\mathbf{\hat{U}}_{i})^{T}\mathbf{\Sigma }^{-1}\},
\end{eqnarray*}%
where $e_{i}=w(\mathbf{\hat{U}}_{i})(\mathbf{\hat{V}}_{i}-\mathbf{\Theta }%
^{T}\mathbf{\hat{U}}_{i})^{T}\mathbf{\Sigma }^{-1}(\mathbf{\hat{V}}_{i}-%
\mathbf{\Theta }^{T}\mathbf{\hat{U}}_{i})$. Then 
\begin{equation}
\nabla _{\mathrm{vec}(\mathbf{\Theta })}m_{(\mathbf{\Theta },\mathbf{\Sigma }%
)}(\mathbf{\hat{U}}_{i},\mathbf{\hat{V}}_{i})=-2\rho ^{\prime }(e_{i})w(%
\mathbf{\hat{U}}_{i})\mathrm{vec}\{\mathbf{\hat{U}}_{i}(\mathbf{\hat{V}}_{i}-%
\mathbf{\Theta }^{T}\mathbf{\hat{U}}_{i})^{T}\mathbf{\Sigma }^{-1}\},
\label{eq:der-wrt-Theta}
\end{equation}%
which can be rearranged in matrix form as 
\[
\frac{\partial }{\partial \mathbf{\Theta }}m_{(\mathbf{\Theta },\mathbf{%
\Sigma })}(\mathbf{\hat{U}}_{i},\mathbf{\hat{V}}_{i})=-2\rho ^{\prime
}(e_{i})w(\mathbf{\hat{U}}_{i})\mathbf{\hat{U}}_{i}(\mathbf{\hat{V}}_{i}-%
\mathbf{\Theta }^{T}\mathbf{\hat{U}}_{i})^{T}\mathbf{\Sigma }^{-1}, 
\]%
and (\ref{eq:fixed-point-Theta}) follows. Differentiating $m$ with respect
to $\mathbf{\Sigma }$ we obtain 
\[
\mathrm{d}m_{(\mathbf{\Theta },\mathbf{\Sigma })}(\mathbf{\hat{U}}_{i},%
\mathbf{\hat{V}}_{i})= 
\]%
\begin{eqnarray*}
&=&\rho ^{\prime }(e_{i})w(\mathbf{\hat{U}}_{i})(\mathbf{\hat{V}}_{i}-%
\mathbf{\Theta }^{T}\mathbf{\hat{U}}_{i})^{T}\{-\mathbf{\Sigma }^{-1}(%
\mathrm{d}\mathbf{\Sigma })\mathbf{\Sigma }^{-1}\}(\mathbf{\hat{V}}_{i}-%
\mathbf{\Theta }^{T}\mathbf{\hat{U}}_{i})+\mathrm{tr}\{\mathbf{\Sigma }^{-1}(%
\mathrm{d}\mathbf{\Sigma })\} \\
&=&-\rho ^{\prime }(e_{i})w(\mathbf{\hat{U}}_{i})\mathrm{tr}\{\mathbf{\Sigma 
}^{-1}(\mathbf{\hat{V}}_{i}-\mathbf{\Theta }^{T}\mathbf{\hat{U}}_{i})(%
\mathbf{\hat{V}}_{i}-\mathbf{\Theta }^{T}\mathbf{\hat{U}}_{i})^{T}\mathbf{%
\Sigma }^{-1}(\mathrm{d}\mathbf{\Sigma })\}+\mathrm{tr}\{\mathbf{\Sigma }%
^{-1}(\mathrm{d}\mathbf{\Sigma })\} \\
&=&-\rho ^{\prime }(e_{i})w(\mathbf{\hat{U}}_{i})\mathrm{vec}\{\mathbf{%
\Sigma }^{-1}(\mathbf{\hat{V}}_{i}-\mathbf{\Theta }^{T}\mathbf{\hat{U}}_{i})(%
\mathbf{\hat{V}}_{i}-\mathbf{\Theta }^{T}\mathbf{\hat{U}}_{i})^{T}\mathbf{%
\Sigma }^{-1}\}^{T}\mathrm{vec}(\mathrm{d}\mathbf{\Sigma }) \\
&&+\mathrm{vec}(\mathbf{\Sigma }^{-1})^{T}\mathrm{vec}(\mathrm{d}\mathbf{%
\Sigma }),
\end{eqnarray*}%
so 
\[
\nabla _{\mathrm{vec}(\mathbf{\Sigma })}m_{(\mathbf{\Theta },\mathbf{\Sigma }%
)}(\mathbf{\hat{U}}_{i},\mathbf{\hat{V}}_{i}) 
\]%
\[
=-\rho ^{\prime }(e_{i})w(\mathbf{\hat{U}}_{i})\mathrm{vec}\{\mathbf{\Sigma }%
^{-1}(\mathbf{\hat{V}}_{i}-\mathbf{\Theta }^{T}\mathbf{\hat{U}}_{i})(\mathbf{%
\hat{V}}_{i}-\mathbf{\Theta }^{T}\mathbf{\hat{U}}_{i})^{T}\mathbf{\Sigma }%
^{-1}\}+\mathrm{vec}(\mathbf{\Sigma }^{-1}). 
\]%
Again, this can be expressed in matrix form as 
\[
\frac{\partial }{\partial \mathbf{\Sigma }}m_{(\mathbf{\Theta },\mathbf{%
\Sigma })}(\mathbf{\hat{U}}_{i},\mathbf{\hat{V}}_{i}) 
\]%
\[
=-\rho ^{\prime }(e_{i})w(\mathbf{\hat{U}}_{i})\mathbf{\Sigma }^{-1}(\mathbf{%
\hat{V}}_{i}-\mathbf{\Theta }^{T}\mathbf{\hat{U}}_{i})(\mathbf{\hat{V}}_{i}-%
\mathbf{\Theta }^{T}\mathbf{\hat{U}}_{i})^{T}\mathbf{\Sigma }^{-1}+\mathbf{%
\Sigma }^{-1}, 
\]%
from which (\ref{eq:fixed-point-Sigma}) follows.

We will simplify the derivation of the asymptotic distribution of $\mathbf{%
\hat{\Theta}}$ by assuming that the true component scores $(\mathbf{U}_{i},%
\mathbf{V}_{i})$ are used, instead of the estimated scores $(\mathbf{\hat{U}}%
_{i},\mathbf{\hat{V}}_{i})$, and by assuming that $\mathbf{\Sigma }_{0}$ is
fixed and known. In that case we can apply Theorem 5.23 of Van der Vaart
(1998) directly, and obtain that $\sqrt{n}\{\mathrm{vec}(\mathbf{\hat{\Theta}%
)-}\mathrm{vec}(\mathbf{\Theta }_{0}\mathbf{)}\}$ is asymptotically $N(%
\mathbf{0},\mathbf{A}^{-1}\mathbf{BA}^{-1})$ with 
\[
\mathbf{A}=\mathrm{E}\{\nabla _{\mathrm{vec}(\mathbf{\Theta })}\nabla _{%
\mathrm{vec}(\mathbf{\Theta })}^{T}m_{(\mathbf{\Theta }_{0},\mathbf{\Sigma }%
_{0})}(\mathbf{U},\mathbf{V})\} 
\]%
and 
\[
\mathbf{B}=\mathrm{E}\{\nabla _{\mathrm{vec}(\mathbf{\Theta })}m_{(\mathbf{%
\Theta }_{0},\mathbf{\Sigma }_{0})}(\mathbf{U},\mathbf{V})\nabla _{\mathrm{%
vec}(\mathbf{\Theta })}^{T}m_{(\mathbf{\Theta }_{0},\mathbf{\Sigma }_{0})}(%
\mathbf{U},\mathbf{V})\}; 
\]%
these expectations are taken with respect to the true parameters $(\mathbf{%
\Theta }_{0},\mathbf{\Sigma }_{0})$. Without loss of generality we can
eliminate the factor $2\mathbf{\Sigma }^{-1}$ in (\ref{eq:der-wrt-Theta});
then it is easy to see that (\ref{eq:B}) holds. To derive (\ref{eq:A}) we
use differentials again: 
\[
\mathrm{d}\{\nabla _{\mathrm{vec}(\mathbf{\Theta })}^{T}m_{(\mathbf{\Theta },%
\mathbf{\Sigma }_{0})}(\mathbf{U},\mathbf{V})\}= 
\]%
\begin{eqnarray*}
&=&2\rho ^{\prime \prime }(e)w^{2}(\mathbf{U})(\mathbf{V}-\mathbf{\Theta }%
^{T}\mathbf{U})^{T}\mathbf{\Sigma }_{0}^{-1}(\mathrm{d}\mathbf{\Theta })^{T}%
\mathbf{U}\mathrm{vec}\{\mathbf{U}(\mathbf{V}-\mathbf{\Theta }^{T}\mathbf{U}%
)^{T}\}^{T} \\
&&+\rho ^{\prime }(e)w(\mathbf{U})\mathrm{vec}\{\mathbf{U}(\mathrm{d}\mathbf{%
\Theta }^{T}\mathbf{U})^{T}\}^{T} \\
&=&2\rho ^{\prime \prime }(e)w^{2}(\mathbf{U})\mathrm{tr}\{(\mathrm{d}%
\mathbf{\Theta })^{T}\mathbf{U}(\mathbf{V}-\mathbf{\Theta }^{T}\mathbf{U}%
)^{T}\mathbf{\Sigma }_{0}^{-1}\}\mathrm{vec}\{\mathbf{U}(\mathbf{V}-\mathbf{%
\Theta }^{T}\mathbf{U})^{T}\}^{T} \\
&&+\rho ^{\prime }(e)w(\mathbf{U})\mathrm{vec}(\mathbf{UU}^{T}\mathrm{d}%
\mathbf{\Theta })^{T} \\
&=&2\rho ^{\prime \prime }(e)w^{2}(\mathbf{U})\mathrm{vec}(\mathrm{d}\mathbf{%
\Theta })^{T}\mathrm{vec}\{\mathbf{U}(\mathbf{V}-\mathbf{\Theta }^{T}\mathbf{%
U})^{T}\mathbf{\Sigma }_{0}^{-1}\}\mathrm{vec}\{\mathbf{U}(\mathbf{V}-%
\mathbf{\Theta }^{T}\mathbf{U})^{T}\}^{T} \\
&&+\rho ^{\prime }(e)w(\mathbf{U})\{(\mathbf{I}_{q}\otimes \mathbf{UU}^{T})%
\mathrm{vec}(\mathrm{d}\mathbf{\Theta })\}^{T},
\end{eqnarray*}%
so 
\[
\nabla _{\mathrm{vec}(\mathbf{\Theta })}\nabla _{\mathrm{vec}(\mathbf{\Theta 
})}^{T}m_{(\mathbf{\Theta },\mathbf{\Sigma }_{0})}(\mathbf{U},\mathbf{V})= 
\]%
\begin{eqnarray*}
&=&2\rho ^{\prime \prime }(e)w^{2}(\mathbf{U})\mathrm{vec}\{\mathbf{U}(%
\mathbf{V}-\mathbf{\Theta }^{T}\mathbf{U})^{T}\mathbf{\Sigma }_{0}^{-1}\}%
\mathrm{vec}\{\mathbf{U}(\mathbf{V}-\mathbf{\Theta }^{T}\mathbf{U})^{T}\}^{T}
\\
&&+\rho ^{\prime }(e)w(\mathbf{U})(\mathbf{I}_{q}\otimes \mathbf{UU}^{T}) \\
&=&2\rho ^{\prime \prime }(e)w^{2}(\mathbf{U})(\mathbf{\Sigma }_{0}^{-1}%
\mathbf{R}\otimes \mathbf{U})(\mathbf{R}\otimes \mathbf{U})^{T}+\rho
^{\prime }(e)w(\mathbf{U})(\mathbf{I}_{q}\otimes \mathbf{UU}^{T}),
\end{eqnarray*}%
from which (\ref{eq:A}) follows.

\section*{References}

\begin{description}
\item Ash, R. B. and Gardner, M. F. (1975). \emph{Topics in Stochastic
Processes}. Probability and Mathematical Statistics (Vol.~27). New York:
Academic Press.

\item Chiou, J.-M., M\"{u}ller, H.-G., and Wang, J.-L. (2004). Functional
response models. \emph{Statistica Sinica} \textbf{14}, 675--693.

\item Cuevas, A., Febrero, M., and Fraiman, R. (2007). Robust estimation and
classification for functional data via projection-based depth notions.\ 
\emph{Computational Statistics} \textbf{22}, 481--496.

\item Davison, A. C. and Hinkley, D. V. (1997). \emph{Bootstrap Methods and
Their Application.} Cambridge: Cambridge University Press.

\item Fraiman, R., and Muniz, G. (2001). Trimmed means for functional data. 
\emph{Test }\textbf{10}, 419--440.

\item Gervini, D. (2008). Robust functional estimation using the median and
spherical principal components. \emph{Biometrika} \textbf{95}, 587--600.

\item Gervini, D. (2009). Detecting and handling outlying trajectories in
irregularly sampled functional datasets. \emph{The Annals of Applied
Statistics} \textbf{3}, 1758--1775.

\item He, X., Simpson, D. G. and Wang, G. (2000). Breakdown points of $t$%
-type regression estimators. \emph{Biometrika} \textbf{87}, 675--687.

\item James, G., Hastie, T. G. and Sugar, C. A. (2000). Principal component
models for sparse functional data.\ \emph{Biometrika} \textbf{87}, 587--602.

\item Locantore, N., Marron, J. S., Simpson, D. G., Tripoli, N., Zhang, J.
T. and Cohen, K. L. (1999). Robust principal components for functional data
(with discussion).\ \emph{Test} \textbf{8}, 1--28.

\item Magnus, J. R., and Neudecker, H. (1999). \emph{Matrix Differential
Calculus with Applications in Statistics and Econometrics. Revised Edition},
New York: Wiley.

\item Maronna, R. A., Martin, R. D. and Yohai, V. J. (2006). \emph{Robust
Statistics. Theory and Methods}. New York: Wiley.

\item Maronna, R. A. and Yohai, V. J. (2012). Robust functional linear
regression based on splines. To appear in \emph{Computational Statistics \&
Data Analysis.}

\item M\"{u}ller, H.-G., Chiou, J.-M., and Leng, X. (2008). Inferring gene
expression dynamics via functional regression analysis. \emph{BMC
Bioinformatics} \textbf{9:60}.

\item Ramsay, J. O. and Silverman, B. W. (2005). \emph{Functional Data
Analysis. Second Edition}. New York: Springer.

\item Van der Vaart, A. W. (1998). \emph{Asymptotic Statistics}. Cambridge
Series in Statistical and Probabilistic Mathematics, Cambridge, UK:
Cambridge University Press.

\item Yao, F., M\"{u}ller, H.-G. and Wang, J.-L. (2005a). Functional linear
regression analysis for longitudinal data. \emph{The Annals of Statistics }%
\textbf{33}, 2873--2903.

\item Yao, F., M\"{u}ller, H.-G. and Wang, J.-L. (2005b). Functional data
analysis for sparse longitudinal data.\ \emph{Journal of the American
Statistical Association} \textbf{100}, 577--590.

\item Zhu, H., Brown, P.J.~and Morris, J.S. (2011). Robust, adaptive
functional regression in functional mixed model framework. \emph{Journal of
the American Statistical Association} \textbf{106}, 1167--1179.
\end{description}

\end{document}